\documentclass[structabstract]{raa}
\usepackage{graphicx,times}
\usepackage{natbib,amssymb}
\usepackage{epsfig}
\usepackage{float}
\providecommand{\bm}[1]{\mbox{\boldmath$#1$\unboldmath}}

\begin{document}

    \title{Past and future of the central double-degenerate core of Henize 2-428}

	\volnopage{ {\bf 2018} Vol.\ {\bf X} No. {\bf XX}, 000--000}
	\setcounter{page}{1}

	\author{Dong-Hao Wu\inst{1,2,3,4}
		\and
		Dong-Dong Liu\inst{1,2,3,4}
		\and
		Bo Wang \inst{1,2,3,4}}

	\institute{Yunnan Observatories, Chinese Academy of Sciences, Kunming 650216, China；
		{\it wdh@ynao.ac.cn; liudongdong@ynao.ac.cn; wangbo@ynao.ac.cn}\\
		\and
		Key Laboratory for the Structure and Evolution of Celestial Objects, Chinese Academy of Sciences, Kunming 650216, China
		\and
		University of Chinese Academy of Sciences, Beijing 100049, China\\
		\and
		Center for Astronomical Mega-Science, Chinese Academy of Sciences, Beijing 100012, China}

	\date{Received ; accepted}

	\abstract {It has been suggested that SNe Ia could be produced in the condition of the violent merger scenario of the double-degenerate model, in which a thermonuclear explosion could be produced when the merging of double carbon-oxygen white dwarfs (CO WDs) is still ongoing. It has been recently found that the nucleus of the bipolar planetary nebula Henize 2-428 consists of double CO WDs that have a total mass of $\sim$$1.76\, M_{\odot}$, a mass ratio of $\sim$$1$ and an orbital period of $\sim$$4.2\,\rm hours$, which is the first and only discovered progenitor candidate of SNe Ia predicted by the violent merger scenario. In this work, we aim to reproduce the evolutionary history of the central double CO WDs of Henize 2-428. 
We find that the planetary nebula Henize 2-428 may originate from a primordial binary that have a $\sim$$5.4\, M_{\odot}$ primary and a $\sim$$2.7\, M_{\odot}$ secondary with an initial orbital period of $\sim$$15.9\,\rm days$. The double CO WDs are formed after the primordial binary experiencing two Roche-lobe overflows and two common-envelope ejection processes. According to our calculations, it takes about $\sim$$840\,\rm Myr$ for the double CO WDs to merge and form an SN Ia driven by the gravitational wave radiation after their birth. To produce the current status of Henize 2-428, a large common-envelope parameter is needed. 
We also estimate that the rate of SNe Ia from the violent merger scenario is at most $2.9\times 10^{-4}\,\rm yr^{-1}$, and that the delay time is in the range of $\sim$$90\,\rm Myr$ to the Hubble time.
 }

	\titlerunning{Past and future of the central DD core of Henize 2-428}

	\authorrunning{D.-H. Wu et al.}	

	\maketitle

\section{Introduction}
Type Ia supernovae (SNe Ia) are one of the most luminous phenomena in the Universe. They have been used as the standard candle for measuring cosmological distance. By employing correlation between the maximum luminosity and the light curve width of SNe Ia (e.g., Phillips 1993), it has been found that the expansion of the Universe is accelerating (e.g., Riess et al. 1998; Perlmutter et al. 1999). SNe Ia may also have a great influence on the chemical evolution of their host galaxies due to the production of iron-peak elements during SN Ia explosions (e.g., Greegio \& Renzzini 1983; Matteucci \& Greggio 1986; Li et al. 2018). In addition, cosmic rays may be accelerated by SN remnants (e.g., Fang \& Zhang 2012; Yang et al. 2015; Zhou \& Vink 2018). However, the progenitor models for SNe Ia are still under discussion, which may influence the accuracy for measuring the cosmological distance (e.g., Podsiadlowski et al. 2008; Howell 2011; Liu et al. 2012; Wang \& Han 2012; Wang et al. 2013; Maoz et al. 2014; Wang 2018).

It has been suggested that SNe Ia are the thermonuclear explosions of carbon-oxygen white dwarfs (CO WDs) in close binaries (e.g., Hoyle \& Fowler 1960). There are two kinds of competing progenitor models of SNe Ia discussed frequently, i.e., the single-degenerate (SD) model and the double-degenerate (DD) model. In the SD model, a WD accretes material from a non-degenerate companion and explodes as an SN Ia when its mass approach to the Chandrasekar mass limit (e.g. Whelan \& Iben, 1973; Nomoto 1984). In the SD model, the companion could be a main sequence (MS) star, a red giant branch (RGB) star, or a helium (He) star (e.g., Li \& van den Heuvel 1997; Langer et al. 2000; Han \& Podsiadlowski 2004, 2006; Wang et al. 2009a; Ablimit et al. 2014; Wu et al. 2016; Liu et al. 2017a). In the DD model, SNe Ia are arised from the merging of double CO WDs that have a total mass larger than the Chandrasekar mass limit (e.g., Webbink 1984; Iben \& Tutukov 1984), though some studies argued that double WD mergers with sub-Chandrasekar mass may also produce SNe Ia (e.g., Ji et al. 2013; Liu et al. 2017b). Comparing with the SD model, the rate of SNe predicted by DD model is high enough to satisfy observational results (e.g., Yungelson et al. 1994; Han 1998; Nelemans et al. 2001; Ruiter et al. 2009; Liu et al. 2018). The delay time of an SN Ia is defined as the time interval between the formation moment of the primordial binary to the moment when the SN Ia is formed.
The delay time distributions (DTDs) predicted by the DD model roughly follow a single power law, which is similar to that derived by observations (e.g., Maoz et al. 2011; Ruiter et al. 2009; Mennekens et al. 2010; Yungelson \& Kuranov 2017; Liu et al. 2018).
However, some studies show that the merger of double CO WDs may produce accretion induced collapse supernovae and eventually form a neutron star (Nomoto \& Iben 1985; Saio \& Nomoto 1985; Timmes et al. 1994).

It has been proposed that an instaneous explosion could be triggered while the merging process of double CO WDs is still ongoing, leading to the formation of an SN Ia (see Pakmor et al. 2010, 2011, 2012). This is a subclass of the DD model named as the violent merger scenario. Pakmor et al. (2010) found that the violent mergers of two $0.9\, M_{\odot}$ CO WDs may produce 1991bg-like events. Pakmor et al. (2011) suggested that the critical minimum mass ratio of double CO WDs for producing SNe Ia is $0.8\, M_{\odot}$ on the basic of the violent merger scenario. R\"opke et al. (2012) found that the violent merger scenario could also explain the observational properties of SN 2011fe. Sato et al. (2016) simulated a large sample of double CO WDs, and found that the critical minimum mass of each WD for producing SNe Ia is $0.8\,  M_{\odot}$ based on the violent merger scenario. Liu et al. (2016) systematically investigated the violent merger scenario by considering the WD$+$He subgiant channel for the formation of double massive WDs, and found that the WD$+$He subgiant channel may contribute to about 10\% of all SNe Ia in the Galaxy based on the violent merger scenario.

Henize 2-428, a bipolar planetary nebula (PN G049.4+02.4), is $\sim$$1.4\pm0.4\,\rm kpc$ from the solar system (see Santander-Garc\'{i}a et al. 2015). By assuming that the double He II 541.2 nm line profile is caused by the absorption of binaries, Santander-Garc\'{i}a et al. (2015) analysed the light curves of Henize 2-428, and found that its nucleus consists of double nearly-equal-mass CO WDs. The total mass of this system is $\sim$$1.76 \, M_{\odot}$ and the orbital period is $\sim$$4.2\,\rm hours$. According to the violent merger scenario, the double degenerate cores of Henize 2-428 is a strong progenitor candidate of SNe Ia. However, the formation path to the nucleus of Henize 2-428 is still unknown.

In this work, we aim to investigate the evolutionary history of the bipolar planetary nebula Henize 2-428, and provide the rates and DTDs of SNe Ia from the violent merger scenario. In Sect. 2, we introduce our numerical methods. We give the results and discussion in Sect. 3. At last, we provide a summary in Sect. 4.
	
\section{Numerical Methods}

\subsection{Violent Merger Criteria}
The merging of double WDs could trigger prompt detonation and produce SNe Ia under certain conditions. In this work, we assumed that the criteria for violent WD mergers are as follows:
	
(1) The mass ratio of double CO WDs ($q=M_{\rm WD2}/M_{\rm WD1}$) should larger than 0.8, where $M_{\rm WD1}$ is the mass of massive WD, and $M_{\rm WD2}$ the mass of the less-massive one (see Pakmor 2011; Liu et al. 2016).
	
(2) The critical minimum mass of each WD is assumed to be $0.8\, M_{\odot}$ (Sato et al. 2016).
	
(3) The delay times of SNe Ia should be less than the Hubble time, i.e., $t=t_{\rm evol}+t_{\rm GW} \leq t_{\rm Hubble}$, where $t_{\rm evol}$ is the evolutionary timescale from primordial binaries to the formation of double CO WDs, and $t_{\rm GW}$ is the timescale during which double WDs are brought together by gravitational wave radiation, written as:
\begin{equation}
{t_{\rm GW}} = 8 \times {10^7} \times \frac{{{{({M_{\rm WD1}} + {M_{\rm WD2}})}^{1/3}}}}{{{M_{\rm WD1}}{M_{\rm WD2}}}}{P^{8/3}},
\end{equation}
in which $t_{\rm GW}$ is in unit of years, $P$ is the orbital period of double WDs in hours, $M_{\rm WD1}$ and $M_{\rm WD2}$ are in unit of $M_{\odot}$.

By adopting these criteria, we obtained a large number of double CO WD systems that may merge violently and then explode as SNe Ia. Subsequently, we provide the evolutionary path of the double WDs closest to current parameters of Henize 2-428 to approximately provide the evolutionary history of Henize 2-428 and speculate its fate.
	
\subsection{BPS approches}
By employing the rapid binary evolutionary code (Hurley et al. 2000, 2002), we performed a series of Monte Carlo BPS simulations evolving primordial binaries to the merging of double CO WDs. In each simulation, $2\times10^{\rm 7}$ primordial binaries are calculated. The initial parameters and basic assumptions in our Monte Carlo BPS computations listed below are adopted:

(1) The initial metallicity in our simulations is set to be 0.02.
	
(2) We assumed that all stars are in binaries with circular orbits.

(3) The initial mass function from Miller \& Scalo (1979) is adopted for the primordial primaries.
	
(4) The initial mass ratios ($q^{\rm '}=M_{\rm 2}/M_{\rm 1}$) are assumed to distribute uniformly (e.g., Mazeh et al. 1992; Goldberg \& Mazeh 1994), i.e., $n(q^{\rm '})=1$, in which $0\leq q^{\rm '} \leq1$.
	
(5) The distribution of initial separation $a$ is assumed to be constant in $\rm log(a)$ for wide binaries and fall smoothly for close binaries (e.g., Han et al. 1995).
	
(6) The star formation rate is assumed to be constant ($5\, M_{\odot}\rm yr^{\rm -1}$) to approximate the Galaxy over the past $15\,\rm Gyr$ (see Yungelson \& Livio 1998; Willems \& Kolb 2004; Han \& Podsiadlowski 2004), or modeled as a delta function (a single star burst of $10^{\rm 10}\, M_{\odot}$ in stars) to roughly describe elliptical galaxies.
	
\subsection{Common-Envelope Computation}
The common-envelope (CE) evolution play a critical role in the formation of double WDs. However, the prescription for calculating CE ejection is still under debate (e.g. Ivanova et al. 2013). In this work, we adopt the standard energy perspective to simulate the CE ejection process, (see Webbink 1984), written as:
\begin{equation}
\alpha_{\rm CE}(\frac{GM^{\rm f}_{\rm don}M^{\rm f}_{\rm acc}}{2{a_{\rm f}}}-\frac{GM^{\rm i}_{\rm don}M^{\rm i}_{\rm acc}}{2{a_{\rm i}}})= \frac{GM^{\rm i}_{\rm don}M_{\rm env}}{\lambda R_{\rm don}},
\end{equation}
in which $G$, $M_{\rm don}$, $M_{\rm acc}$, $a$, $M_{\rm env}$ and $R_{\rm don}$ are the gravitational constant, the donor mass, the accretor mass, the orbital separation, the mass of the donor's envelope and the donor radius, respectively. The superscripts i and f stand for these values before and after the CE ejection.
From this prescription, we can see that there are two variable parameters, i.e. the CE ejection efficiency ($\alpha_{\rm CE}$) and a stellar structure parameter ($\lambda$).
These two parameters may change with the evolutionary process (e.g., Ablimit et al. 2016). It has been suggested that the value of $\alpha_{\rm CE}$ may vary with WD mass, secondary mass, mass ratio, or orbital period (e.g., de Marco et al. 2011; Davis et al. 2012). Meanwhile, the values of $\lambda$ could be investigated by considering gravitational energy only, adding internal energy, or adding the entropy of the envelope (e.g., Davis et al. 2010; Xu \& Li 2010). However, the value of $\alpha_{\rm CE}$ and $\lambda$ are still highly uncertain. In the present work, similar to our previous studies (e.g., Wang et al. 2009b), we simply combined these two parameters as a single free one (i.e. $\alpha_{\rm CE}\lambda$) based on Eq.\,(2), and assumed $\alpha_{\rm CE}\lambda=1$, 2 and 3 to check its effect on the final results.

\section{Results and Discussions}

\begin{figure}
	\begin{center}
		\includegraphics[width=9cm,angle=0]{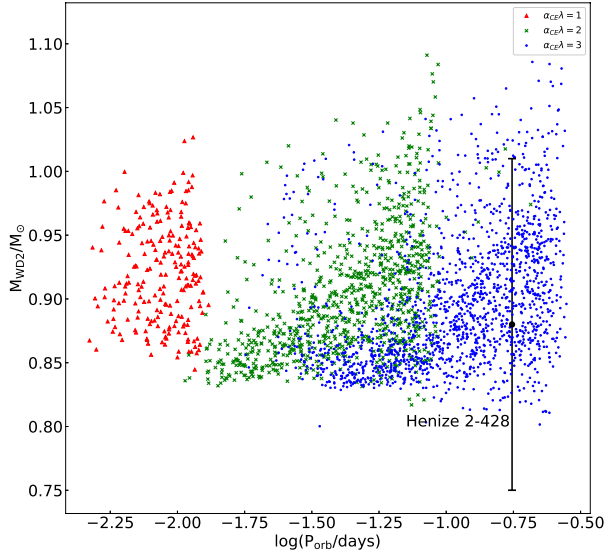}
		\caption{The distribution of violent WD mergers that can produce SNe Ia in the orbital period$-$secondary mass ($\log P_{\rm orb}-M_{\rm WD2}$) plane. Red triangles, green crosses and blue dots represent the simulated results with $\alpha_{\rm CE} \lambda=1$, 2 and 3, respectively. The filled circle with error bar represent the position of the central double degenerate cores of Henize 2-428 (Santander-Garc\'{i}a et al. 2015).}
	\end{center}
\end{figure}

\begin{figure*}
	\begin{center}
		\includegraphics[width=12.5cm,angle=0]{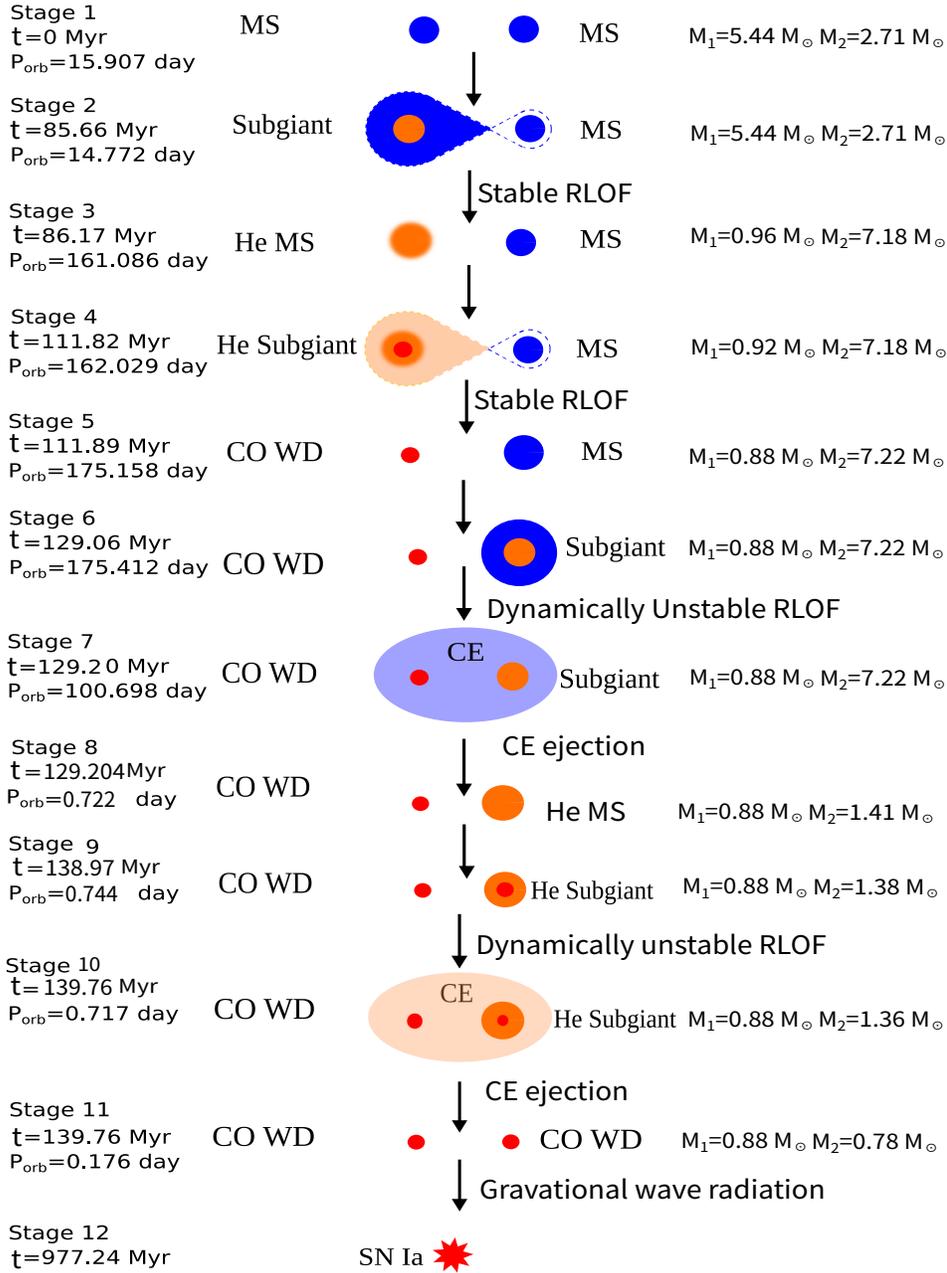}
		\caption{The evolutionary history and future of the planetary nebula Henize 2-428.}
	\end{center}
\end{figure*}

\begin{figure*}
	\begin{center}
	\includegraphics[width=9cm,angle=0]{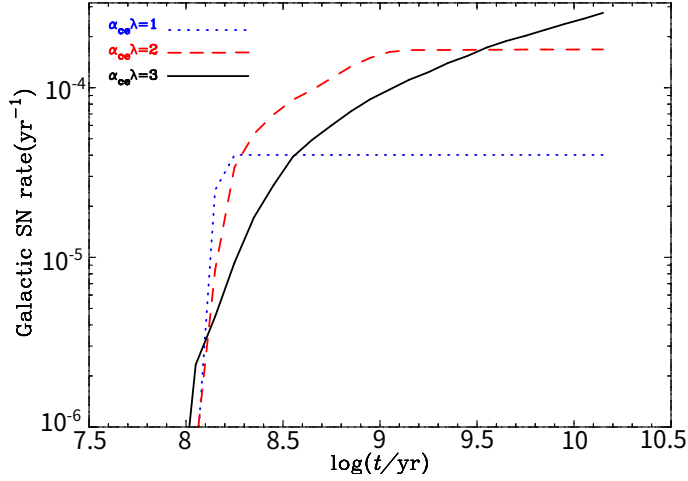}
	\caption{Evolution of SN Ia rates in the Galaxy based on the violent merger scenario. Here, we adopt a constant star formation rate of $5\,M_{\odot} \rm yr^{-1}$. The blue dotted, red dashed and black solid curves corresponds to the cases with $\alpha_{\rm CE} \lambda=1$, 2 and 3, respectively.}
	\end{center}
\end{figure*}
	
\begin{figure*}
	\begin{center}
	\includegraphics[width=9cm,angle=0]{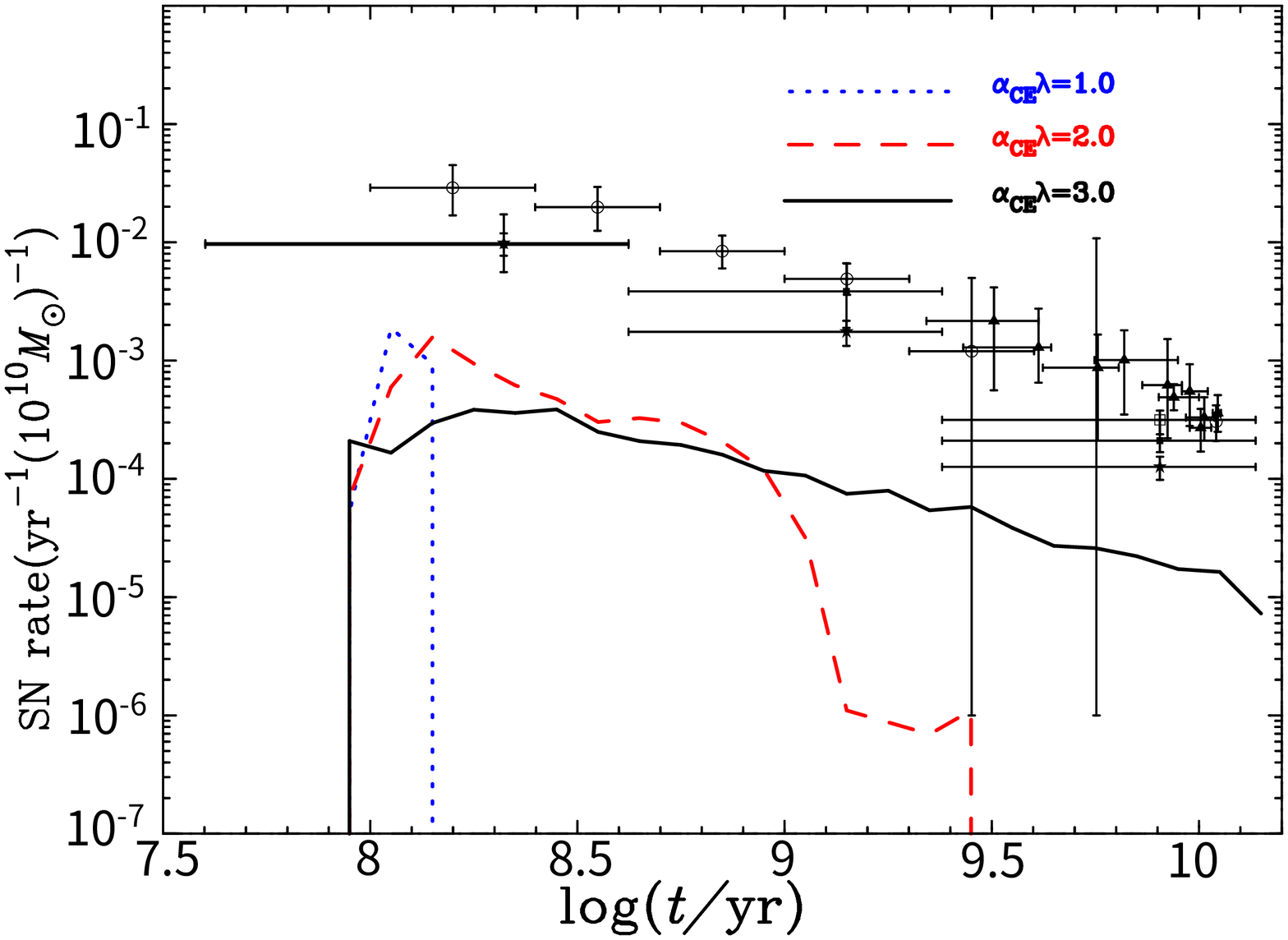}
	\caption{Simalr to Fig.\,3, but for the delay time distributions of SNe Ia. Here, we adopt a star burst of $10^{10}\, M_{\odot}$ in stars. The open circles are from Totani et al. (2008), the filled triangles, squares are taken from Maoz et al. (2010, 2012) and the open square is from Graur \& Maoz (2013).}
	\end{center}
\end{figure*}

Fig.\,1 presents the distribution of double CO WDs that can produce SNe Ia via the violent merger scenario in the $\log P_{\rm orb}-M_{\rm WD2}$ plane. We find that as the value of $\alpha_{\rm CE}\lambda$ increases, the distribution of orbital periods becomes wider and more double WDs would be produced. The reason is that, a larger value of $\alpha_{CE}\lambda$ means that the less amount of orbital energy would be used for unbinding CE and the ejection of CE would be more easily to occur.  The nucleus of Henize 2-428 consists of two nearly-equal-mass $0.88\pm0.13\, M_{\odot}$ CO WDs, and their orbital period is $\sim$$4.2\,\rm hours$ (Santander-Garc\'{i}a et al. 2015). In this figure, we also show the position of the central double degenerate nucleus of Henize 2-428.
Note that for the case of $\alpha_{\rm CE}\lambda=3$, the parameters of several formed double WDs falls within the range of the observation error of Henize 2-428. We adopted the double WDs that is closest to current parameters of Henize 2-428 to approximate the evolutionary path of Henize 2-428 (see Fig.\,2).

Fig.\,2 shows the evolutionary path of Henize 2-428. The primordial binary consists of a $5.4\, M_{\odot}$ primary and a $2.7\, M_{\odot}$ secondary, with an initial orbital period of $\sim$$15.9\,\rm days$. The primordial primary evolved to a subgiant after about $85.47\,\rm Myr$. The radius of primary reaches $22.39\,\rm R_{\odot}$ and fills its Roche-lobe at $t=85.67\,\rm Myr$ (Stage 2), resulting in a stable mass-transfer process.
After about $0.5\,\rm Myr$, the H-rich shell of the primordial primary is exhausted and the mass-transfer stops. In this case, the primordial primary becomes a $0.97 \, M_{\odot}$ He MS star, and the primordial secondary turns to a $7.18\, M_{\odot}$ MS star (Stage 3). The orbital period at this stage is $\sim$$162\,\rm days$.
At $t=111.82\,\rm Myr$, the primordial primary becomes a He subgiant with a radius of $54.95\,\rm R_{\odot}$ and fills its Roche-lobe again, leading to another stable mass-transfer process (Stage 4).
When He-rich shell of the primordial primary is exhausted, the binary evolve to a $0.88\, M_{\odot}$ CO WD and a $7.22\, M_{\odot}$ MS star with an orbital period of $175\,\rm days$ (Stage 5).
Subsequently, the primordial secondary continues to evolve, and will fill its Roche-lobe when it becomes a subgiant star with a radius of $102.8\,\rm R_{\odot}$ at $t=129.06\,\rm Myr$ (Stage 6).
At this stage, the mass transfer is dynamically unstable, leading to the formation of the first CE (Stage 7).
After CE ejection, the orbital period shrinks to $0.722\,\rm days$, and the primordial secondary becomes a $1.43 \, M_{\odot}$ He star (Stage 8).
The He star continues to evolve, and will fill its Roche-lobe again after it evolves to the He subgiant stage at about $t=140\,\rm Myr$ (Stage 9).
At this stage, a CE would be formed due to the dynamically unstable mass-transfer (Stage 10).
After the CE ejection, the binary becomes a double WDs with the nearly equal mass, in which $M_{\,\rm WD1}=0.88 \, M_{\odot}$ and $M_{\,\rm WD2}=0.78 \, M_{\odot}$. During this process, the orbital period shrinks to $0.716\,\rm days$ (Stage 11). The formed double WDs fits well with the observed parameters of Henize 2-428, i.e., the evolutionary history of Henize 2-428 is reproduced. Previous works on the shape of nebula have revealed that the bipolar nebula originate from the CE ejection process (e.g., Han et al. 1995). According to our calculations, we found that the bipolar planetary nebula Henize 2-428 may evolve from the binary in the CE phase with two CO cores.
Afterwards, the formed double WDs will merge driven by the gravitational wave radiation $838\,\rm Myr$ latter, resulting in the production of an SN Ia via the violent merger scenario at about $t=977\,\rm Myr$ (Stage 12). 

Fig.\,3 shows the evolution of SN Ia Galactic rates based on the violent merger scenario.
In this figure, we adopt a constant star formation rate of $5 \, M_{\odot}\rm yr^{-1}$.
From this figure, we can see that the Galactic rates of SNe Ia range from $0.4\times 10^{-4} \,\rm yr^{\rm -1}$ to $2.9\times 10^{-4} \,\rm yr^{\rm -1}$. In the observations, the Galactic SN Ia rate is about 3$-$$4\times 10^{-3} \,\rm yr^{-1}$, that is, the violent merger scenario may contribute to about 1$-$10\% of all SNe Ia in the Galaxy. Note that the rate increases with the values of $\alpha_{\rm CE} \lambda$. That is because, for the case with a larger value of $\alpha_{\rm CE}\lambda$, more double WD systems would be produced (see Fig.\,1).
Note that Ablimit et al. (2016) also provided that the Galactic SN Ia rate is in the range of $8.2\times10^{\rm -5}\,\rm yr^{\rm -1}$$—$$1.7\times10^{\rm -4}\,\rm yr^{\rm -1}$ based on the violent merger scenario, which is generally similar with the results from the present work.

Fig.\,4 displays the DTDs of SNe Ia predicted by the violent merger scenario.
Here, we adopt a star burst of $10^{10}\, M_{\odot}$ in stars.
The delay times of SNe Ia from the violent merger scenario are in the range of $\sim$$90\,\rm Myr$ to the Hubble timescale, which contributes to SNe Ia with young, intermediate and old ages.
For the cases of $\alpha_{\rm CE}\lambda=1$ and 2, the large end of $\log(t)$ is real cut-off on the basic of our calculations.
For the case of $\alpha_{\rm CE}\lambda=3$, the large end is artificial because the time has already reached the Hubble time.

Note that the likelihood for the formation of double WDs with unite mass ratio is still under debate. García-Berro et al. (2015) argued that it is difficult to produce double WDs that has the unit mass ratio, and that this kind of double WDs is rare. The present work provide a possible path for the formation of double WDs with unit mass ratio, and speculate that the number of double WDs with unit mass ratio may be not negligible, which is consistent with the results of Santander-Garc\'ia et al. (2015) and Ablimit et al. (2016).

For the CE ejection parameters, previous studies on the DD model of SNe Ia usually assumed that the values of $\alpha_{\rm CE} \lambda$ ranging from about 0.5 to 2.0 (e.g., Yungelson \& Kuranov 2017; Liu et al. 2018). However, a larger CE ejection parameter is also widely used. Nelemans et al. (2000) studied the formation of double He WDs and they found the CE parameter $\alpha_{\rm CE}\lambda$ could be in range of 1 to 3. Some observations on the post CE binaries show that the values of $\alpha_{\rm CE} \lambda$ may vary from 0.01 to 5 (e.g., Zorotovic et al. 2010).
In this work, we found that in order to reproduce the current stage of the planetary nebula Henize 2-428, a large CE ejection parameter of $\alpha_{\rm CE} \lambda=3$ is needed.

\section{Summary}
	
In the present work, we reproduced the evolutionary history and predict the future of the planetary nebula Henize 2-428. We found that the planetary nebula may originate from a primordial binary that have a $\sim$$5.4\, M_{\odot}$ primary and a $\sim$$2.7\, M_{\odot}$ secondary with an initial orbital period of $\sim$$15.9\,\rm days$. After the birth of the double CO WDs, they would merge and produce an SN Ia through the violent merger scenario after about $\sim$$840\,\rm Myr$. In order to form Henize 2-428, a large CE parameter ($\alpha_{\rm CE}\lambda=3$) is needed. According to our calculations, we also found that the Galactic rate of SNe Ia are in the range 0.4$-$$2.9\times 10^{-4} \,\rm yr^{-1}$ and the delay times range from $\sim$$90\,\rm Myr$ to the Hubble timescale. For a better understanding on the violent merger scenario of SNe Ia, more numerical simulations and more candidates on the double WDs in the observations are required.

\begin{acknowledgements}
We acknowledge useful comments and suggestions from the anonymous referee.
We acknowledge useful comments and suggestions from Zhanwen Han.
We would like to thank Linying Mi for techniacal support. We appreciate Chinese Astronomical Data Center (CAsDC) and Chinese Virtual Observatory (China-VO) for offering computation paltform.
This study is supported by the National Natural Science Foundation of China (Nos 11873085, 11673059 and 11521303), the Chinese Academy of Sciences (Nos QYZDB-SSW-SYS001 and
KJZD-EW-M06-01), and the Yunnan Province (Nos 2017HC018 and 2018FB005).
\end{acknowledgements}

\label{lastpage}
\end{document}